\documentclass{elsart5p}

\usepackage{graphics}
\usepackage{graphicx}
\usepackage{amssymb}
\begin{document}

\newcommand{\raiser}[1]{\raisebox{\upit}[0cm][0cm]{#1}}
\newcommand{\ltappr}{{{\lower4pthbox{$<$} } \atop \widetilde{ \ \ \ }}}
\newlength{\bxwidth}\bxwidth=1.5 truein
\newcommand\frm[1]{\epsfig{file=#1,width=\bxwidth}}
\newcommand{\cg}{{\cal G}}
\newcommand{\dif}[2]{\frac{\delta #1}{\delta #2}}
\newcommand{\ddif}[2]{\frac{\partial #1}{\partial #2}}
\newcommand{\Dif}[2]{\frac{d #1}{d #2}}
\newcommand{\str}{\hbox{Str}}
\newcommand{\Str}{\underline{\hbox{Str}}}
\newcommand{\tr}{{\hbox{Tr}}}
\newcommand{\Tr}{\underline{\hbox{Tr}}}
\newcommand{\dg}{^{\dagger }}
\newcommand{\vk}{\vec k}
\newcommand{\vq}{{\vec{q}}}
\newcommand{\vp}{\bf{p}}
\newcommand{\al}{\alpha}
\newcommand{\be}{\beta}
\newcommand{\si}{\sigma}
\newcommand{\rarrow}{\rightarrow}
\def\fig#1#2{\includegraphics[height=#1]{#2}}
\def\figx#1#2{\includegraphics[width=#1]{#2}}
\newlength{\figwidth}
\figwidth=10cm
\newlength{\shift}
\shift=-0.2cm
\newcommand{\fg}[3]
{
\begin{figure}[ht]

\vspace*{-0cm}
\[
\includegraphics[width=\figwidth]{#1}
\]
\vskip -0.2cm
\caption{\label{#2}
\small#3
}
\end{figure}}
\newcommand{\fgb}[3]
{
\begin{figure}[b]
\vskip 0.0cm
\begin{equation}\label{}
\includegraphics[width=\figwidth]{#1}
\end{equation}
\vskip -0.2cm
\caption{\label{#2}
\small#3
}
\end{figure}}

\newcommand \bea {\begin{eqnarray} }
\newcommand \eea {\end{eqnarray}}
\newcommand{\bk}{{\bf{k}}}
\newcommand{\bx}{{\bf{x}}}

\begin{frontmatter}



\title{Quantum criticality and the break-up of the Kondo
pseudo-potential}
%

\author[AA]{Eran Lebanon\corauthref{Lebanon}},
\ead{lebanon@phys.huji.ac.il}
\author[BB]{P. Coleman}

\address[AA]{Racah Institute of Physics, The Hebrew University, Jerusalem Israel}
\address[BB]{Center for Materials Theory,
Rutgers University, Piscataway, NJ 08854, U.S.A. }

\corauth[Lebanon]{Corresponding author}

\begin{abstract}
We discuss how Anderson's ideas of nominal and real valence can be
incorporated into the current discussion of heavy electron quantum criticality.
In the heavy electron phase, the nominal valence of a screened magnetic ion
differs from its real valence by one unit.  We identify this
discrepancy with the formation of a positively charged background
we call the Kondo pseudo-potential.  At the quantum critical point,
the sudden collapse of the heavy electron Fermi surface can be
identified with the return of the nominal to the real valence. 
This leads to the interesting idea that the heavy electron quantum
critical point may involve locally critical charge degrees of freedom.
We discuss how this might come about within a large $N$ Schwinger
boson scheme. 
\end{abstract}


\end{frontmatter}



\section{Introduction}


Over the course of the development of ideas in heavy fermion physics,
stretching back to the early seventies,
there have been several shifts in opinion about the
relative importance of charge and spin degrees of freedom in heavy
electron systems. The early
view was that heavy electron behavior is driven by slow valence
fluctuations\cite{varma} .
In the  eighties, the discovery that most heavy electron
systems lie close to integral valence,
led to the community to abandon the idea
mixed valence in favor of Doniach's Kondo lattice concept \cite{doniach}
in which  heavy electrons 
form via the Kondo effect
between neutral, localized moments  and the surrounding conduction electrons.
Yet even in this context,  charge plays an important role, 
for the Kondo effect produces resonant scattering that enlarges the
Fermi surface.
Paradoxically, even though the density of
quasiparticles increases, the average electronic charge density is unchanged by the
Kondo effect.
This paradox has  assumed a new level of importance with
the discovery of quantum criticality in heavy electron systems.
The observation of a jump in the Hall constant \cite{silke} and
the heavy electron Fermi surface \cite{shishido}  at heavy
electron quantum  critical points have led many \cite{RCPZ,LC,senthil,pepin,pepinpaul}
to suggest that charge degrees of freedom may play an
important role in heavy electron quantum criticality.


\figwidth = 0.95\columnwidth
\fg{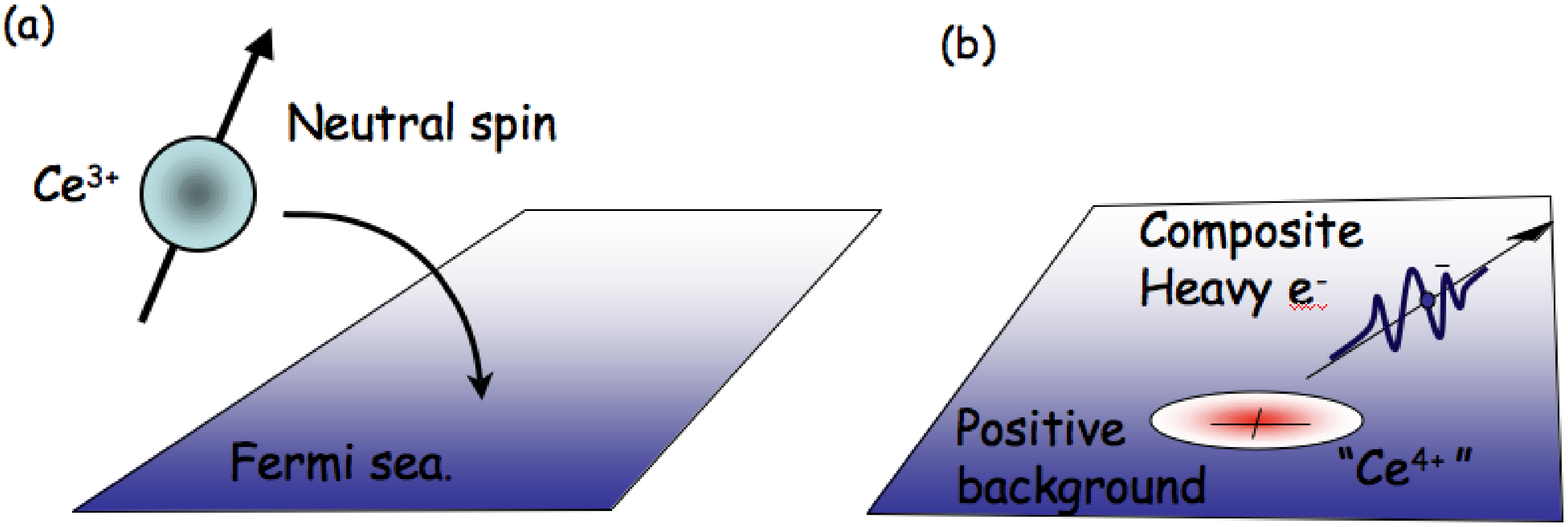}{fig1}{``Ionic'' view of the Kondo effect. Neutral local moment
(a) dissolves in a Fermi sea to (b) produce a negatively charged composite
heavy fermion moving in a positively charged Kondo pseudo-potential.
}

In the early eighties, Anderson introduced the idea of
``nominal valence" to account for the discrepancy between the actual
measured valence of a heavy electron ion, and the ``nominal valence"
required to account for its large Fermi surface
and transport properties\cite{Anderson}. For example, in Cerium heavy
electron metals, such as $CeCu_{6}$,
the $Ce$ ions have a {\sl real} valence of three, with
a localized $4f^1$ local moment within a truly trivalent
$Ce^{3+}$ ion.  
However, when the Kondo effect takes place at low temperatures, each local moment
liberates a mobile charged heavy fermion and the Fermi surface expands
accordingly.  In Anderson's
picture,  $Ce$ ions in heavy electron systems have a {\sl nominal} valence
of four, and behave as spinless $Ce^{4+}$ ions, each donating a
resonant f-electron  to the Fermi sea(see Fig. 1.):
\begin{equation}\label{}
Ce^{3+}\longrightarrow \hbox{Heavy $e^{-}$} + Ce^{4+}.
\end{equation}
This many-body process resembles the dissolution of a neutral atom in a polar
fluid. 

Fisk and Aeppli have emphasized the usefulness of Anderson's idea for
understanding the physics of Kondo insulators\cite{fisk}, where the
``expansion'' of the Fermi surface produces a completely filled band,
leading to a renormalized band insulator.
For example,  the Kondo insulator $Ce_3Bi_4Pt_3$, containing $Ce^{3+}$ ions
is iso-electronic with the band insulator
$Th_{3}Sb_{4}Ni_{3}$, containing $Th^{4+}$ ions.
The discovery of Kondo insulators and the observation of large
Fermi surfaces in heavy electron systems are testament to the
usefulness reality of Andersons concept of nominal and real valence.

\section{The Kondo pseudo-potential}

Embodied in these early ideas, is a notion of elementary charge conservation.
In this paper
we examine the consequences of properly accounting for charge
conservation in the Kondo effect. We argue that 
if a neutral spin liberates a mobile, negatively charged heavy
electron, then it must leave behind a postively charged background.
We call this background
the ``Kondo pseudo-potential''.

The key idea behind this charge book-keeping is embodied in the ``Anderson Clogston''
compensation theorem\cite{clogston61}, which states that the average charge  density
is unchanged by the introduction of a magnetic ion into the Fermi
sea. In the lattice, this 
leads to the following relationship between the large
Fermi surface volume and the conduction electron charge density
\begin{eqnarray}\label{l}
n_{e}= \overbrace{2 \frac{v_{F}}{(2 \pi)^{3}}}^{\hbox{\tiny mobile
HF}} - \overbrace{\frac{Q}{a^{3}}}^{\hbox{\tiny charged }\atop
\hbox{\tiny  background}}
\end{eqnarray}
This kind of book-keeping emerges naturally in both the slave boson
and more recent Schwinger boson descriptions of the Kondo lattice.

However, the
importance of the Kondo pseudo-potential extends beyond simple
book-keeping:  as  a real phenomenon, it has tangible, measurable  conseqeunces.
At the level of a mean-field pseudo-potential, it governs the
symmetries of the heavy electron Fermi surface - symmetries that are
visible not just in the de Haas van Alphen measurements but are
expected to manifest themselves as spatial redistributions in the
local density of states, visible to STM measurements.
More-over, the Kondo pseudo-potential has specific translational,
orbital and gauge symmetries, all of which can, in principle develop
spontaneous broken symmetries, leading to
new phases of heavy electron materials with long-range order:

\vskip 0.2in
\noindent $\bullet $ Broken translation symmetry $\rightarrow$  HF
density wave
\\
$\bullet $ Broken lattice symmetry $\rightarrow$ orbital order
\\
$\bullet $
Broken gauge symmetry $\rightarrow $ HF superconductivity.
\\

\noindent These are all consequences of a rigid, or classical pseudo-potential,
as obtained, for example in a slave boson mean field theory, or a
Korringa Kohn Rostoker (KKR) pseudo-potential approach to the Kondo lattice\cite{Zwicknagl}.

However, the principle focus of this paper is the idea that Kondo pseudo-potential
is not just a rigid  classical background, but a quantum object with
its own set of excitations.
This basic idea forms  a common thread behind many
current efforts to understand heavy electron quantum criticality.
These excitations might be just bosonic fluctuations
of the background pseudo-potential, as described by the Gaussian
fluctuations about a slave boson mean field theory
\cite{senthil, pepinpaul}.
However, our recent calculations based on a Schwinger boson approach
to the Kondo lattice\cite{RCPZ,LRCP} have raised the interesting possibility that the excitations
of the Kondo pseudo-potential might be gapped excitations that carry
either spin or charge.  Closure of the gap at a quantum critical point
leads to a deconfinement of these excitations and the break-down of the Kondo pseudo-potential.

\section{Theoretical realization of the Kondo pseudo potential}

\subsection{Slave Boson Approaches}\label{}

Mean field approaches to the Kondo
lattice\cite{coleman,RN,millis,levin}
based on the slave boson formalism provide an elegant demonstration of the idea of the Kondo
pseudo-potential, with its positiviely charged background.
The mean field Hamiltonian is 
\begin{equation}
{\cal H}_{\rm MF} = {\cal H}_c + \sum_j {\cal H}_I^j(V,\lambda) .
\end{equation}
where ${\cal H}_c$ describes the conduction electrons, and
\begin{equation}\label{}
{\cal H}_I^j = V\sum_{\alpha} ( c^{\dagger}_{j\alpha}
f_{j\alpha} + {\rm  H.c.}) +\lambda (n_{fj}-Q)
\end{equation}
describes the  resonant f-level within
the Kondo pseudo-potential, where $c\dg_{j\alpha } $ and
$f\dg_{j\alpha } $ create the conduction and composite f-electrons, while
$V$ and $\lambda$ are the self-consistently determined hybridization and f-level
position. In the language of field theory, the local $U (1)$ gauge
invariance associated with the neutral Abrikosov fermions is 
broken by the development of the hybridization $V$. 
The associated Higgs effect  ``pins'' the internal gauge field $\lambda$
to the external electrostatic potential, so that 
the f-electrons acquire a  physical charge and the constraint term $Q$
becomes a positive background charge. 

To see this in detail, consider the introduction of  an external electric
potential field $\Phi (\vec{x},t)$ coupled to the conduction
electrons. The electromagnetic gauge invariance is given by
\begin{eqnarray}\label{l}
c_{j\alpha }&\longrightarrow & e^{-i\alpha_{j} (t)} c_{j\alpha },
\qquad e\Phi_{j} (t)\rightarrow e \Phi_{j} (t) - \dot\alpha_j (t)
\cr f_{j\alpha }&\longrightarrow & e^{-i\alpha_{j} (t)} f_{j\alpha
}, \qquad \lambda_{j} (t)\rightarrow \lambda_{j} (t) +\dot\alpha_j
\end{eqnarray}
Notice that the f-electron transforms in the same way as the
conduction electron, implicating its acquisition of charge. The
$\lambda$ field transforms in the same way as the external
potential, implicating its new role as a physical electrostatic
potential.

If we expand the action around its mean-field saddle point in the
presence of a small external potential, it takes the form
\begin{equation}\label{}
S = S_{0} - \sum_{j} \eta \int dt  \frac{(\delta \lambda_{j} (t)+e\Phi
(j,t))^{2}}{2T_{K}}
\end{equation}
where $\eta $ is a constant of order unity and $T_{K}$ is the Kondo
temperature.  The particular combination of $\delta \lambda_{j}$ and
$e\Phi_{j}$ is determined by the electromagnetic gauge invariance.
From the saddle point condition $\delta S/\delta \lambda_{j} (t)=0
\Longrightarrow (\delta  \lambda_{j}-e\Phi_{J})=0$, we see that in a
small electrostatic field, the f-level position picks up a small
time dependent component $\delta \lambda_{j} (t)= - e \Phi_{j} (t)$.
If we now look back at the original Kondo pseudo-potential, we must
replace $\lambda\rightarrow \lambda - e \Phi_{j} (t)$, so that it
becomes
\begin{equation}\label{}
{\cal H}_I^j = V\sum_{\alpha} ( c^{\dagger}_{j\alpha}
f_{j\alpha} + {\rm  H.c.}) + (\lambda - e\Phi_{j} (t)) (n_{fj}-Q)
\end{equation}
From this we not only see that the f-electrons have acquired a
negative charge, but that the external potential has a source term
of the form $+e \Phi_{j}Q$, describing a positively charged
background $+Q$ in the pseudo-potential.

\subsection{Schwinger boson approach}

The Schwinger boson approach to the Kondo lattice
provides a dynamical description of the Kondo pseudo-potential, which
is given by
\begin{equation}\label{}
H_{I} (j) = \sum_{\sigma \in [1,N], \nu\in[1,2S]}\biggl[c\dg_{j\sigma \nu} (\chi \dg_{j\nu}b_{j\sigma }) +
{\rm H. c}\biggr] + \frac{\chi \dg_{j\nu}\chi_{j\nu}}{J}
\end{equation}
where $b_{j\sigma }$ is the ``spinon'' field used in the
Schwinger boson description of the Kondo spin $S_{\alpha \beta } (j) =
b\dg_{j\alpha }b_{j\beta}$ and $\chi_{j\nu}$
is a charged ``holon'' field that mediates the Kondo interaction and carries
the channel index $\nu \in [1,2S]$.  The number of channels is
kept precisely equal to $2S$, so that the spin $S$ is perfectly
screened\cite{RCPZ,LRCP,PG}. In the limit that $N\rightarrow
\infty $ keeping $2S/N$ fixed, the Kondo impurity and Kondo lattice models
can be solved exactly. 

The Schwinger boson approach offers two advantages over its slave
boson partner - for the single impurity case, the large $N$ limit
provides a smooth description of the cross-over from local moment to
local Fermi liquid physics, avoiding the false phase transition  of
the slave boson approach.  More importantly however, this approach
allows the inclusion of magnetism, and can be directly connected with
the Arovas Auerbach\cite{auerbach} approach to quantum magnetism. 

One of the fascinating aspects of the large $N$ solution, is that the
holon and ``spinon'' develop a gap in their excitation spectrum, which
guarantees that the low energy excitations of system are exclusively
electron quasiparticles.  The positively charged holons form a 
filled band, providing the
background charge. 
As the conduction electrons propagate
through the Kondo pseudo-potential, they generate virtual spinon-holon pair
excitations which renormalize their mass and change the Fermi surface
volume (Fig. 2a).
While, in the slave
boson scheme, the Kondo pseudo-potential is described by the 
resonant scattering potential, $\Sigma (\omega) = V^{2}/ (\omega-\lambda)$, in
the Scwhinger boson schem, this is replaced by a dynamical self energy
of the form
\begin{eqnarray}
\Sigma_{c}({\vec r},\tau) =\ \ \ {\cal G}_{\chi}^0(-{\vec
r},-\tau){\cal G}_b({\vec r},\tau) 
\end{eqnarray}
Using diagrammatic means, expanding the Free energy in powers of $1/N$,
it is now possible to extend Luttinger's  original derivation of the Fermi
surface sum rule to the Schwinger boson description of the Kondo
lattice\cite{paul}:
\begin{equation}
\frac{n_{e}}{2S} = N \frac{V_{FS}}{(2\pi)^3} - \frac{V_{\chi}}{2\pi)^3}.
\end{equation}
where $ \frac{V_{\chi}}{2\pi)^3} = 1$, for the filled holon band,
producing the background charge. 

\figwidth = 0.7\columnwidth
\fg{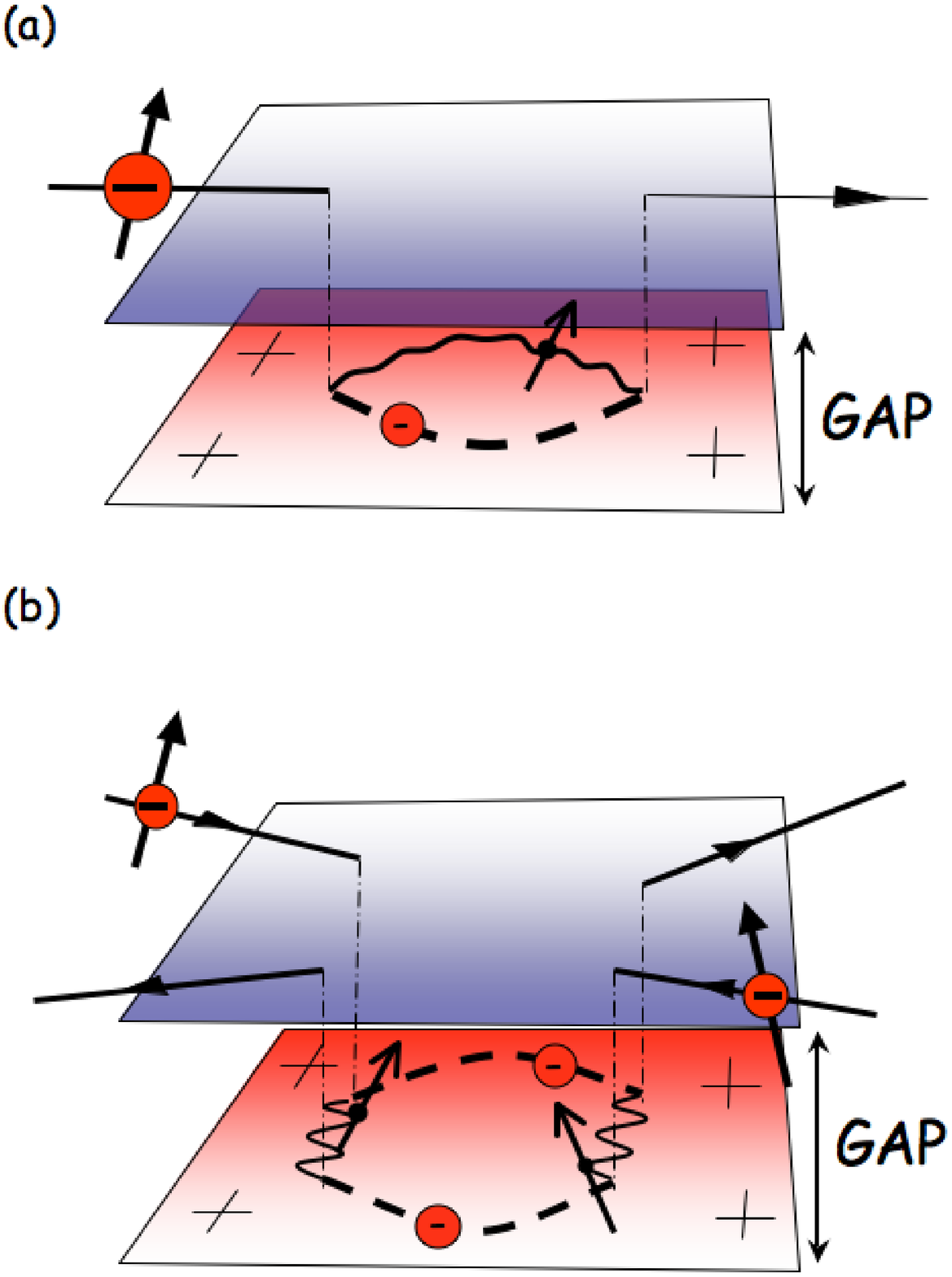}{fig2}{(a) Virtual excitation of gapped spinon/holon pair
renormalizes the heavy electron mass (b) Fermi liquid parameters are
generated by the exchange of gapped spinons and holons. 
}
One of the most important aspects of this theory, is that 
the exchange of virtual spinons and holons mediates the Fermi
liquid interactions (Fig. 2b). In the case of the impurity model, we are
able to derive the analytic properties  of these Fermi liquid
interactions\cite{LC}. For example, we
are able to show that the interaction of the local moment with the
conduction sea preserves the ``Shiba- Korringa'' relation between the
local susceptibility $\chi $ and the spin-spectral weight
\begin{equation}\label{}
\left. \frac{\chi '' (\omega)}{\omega}\right|
_{\omega=0} = \frac{\pi}{2NS}\chi^{2}.
\end{equation}

These ideas are of particular interest in the context of heavy
quantum criticality.  In the Schwinger boson picture,
magnetism is signalled by the condensation of the Schwinger bosons and
provided the phase transition is second order, gauge invariance
guarantees that the spinon and the holon gaps must close together.
It is this process that gives rise to the break-up of the Kondo
pseudo-potential and the development of non Fermi liquid behavior 
in the Schwinger boson scheme, 

\section{Deconfinement and the breakup of the Kondo Pseudo-potential}

Anderson's ideas of nominal and real valence, and the idea of the
Kondo pseudo-potential  acquire an interesting significance
in the context of heavy electron criticality.  If the heavy fermi
surfaces does indeed collapse at a QCP,  then it follows that the
Kondo pseudo-potential must break-up at this point, and the nominal
valence of each Kondo ion must revert to its real value. This is true
in both the Schwinger boson and the slave boson description. 

In the Schwinger boson approach however, heavy electron quantum
criticality is the point where the both the spinon and the holon degrees of freedom become
gapless and deconfine.   We envision the following scenario: as the
strength of the Kondo coupling is reduced, the spinon-holon gap
shrinks to zero and the filled holon band rises to zero energy.
At the quantum critical point the holons annihilate with part of the
heavy electron sea to produce a spin, causing the nominal valence to
revert to the real valence. 

One of the simplest types of solution to seek in the large $N$ limit 
involves the idea of local holon quantum criticality.  This provides an unexpected
explanation of the local critical character of
the spin fluctuations at a heavy electron quantum critical point\cite{schroeder,qmsi}.
In the large $N$ limit, the holon self energy is governed by a
product of the conduction electron and spinon propagators
\[
\Sigma_{\chi } (\bx,\tau)= G^{{(0)}}_c(-\bx,-\tau)G_{b} (\bx,\tau)
\]
where $G_{c} (\bx,\tau )$ is the bare conduction propagator and $G_{b}
(\bx ,\tau )$ the spinon-propagator. The leading long-time singularity of $G^{(0)}_{c}$ is local
\begin{equation}\label{}
G_{c}^{(0)} (\bx,\tau)\sim \frac{\rho }{\tau }\delta (\bx)
\end{equation}
and this means that even though the condensing spinons have a highly
non-local propagator, the leading singular behavior of the holon self
energy and propagator $G_{\chi }$  will  also be local. This suggests
that  that at criticality,  the break-up of the Kondo pseudo-potential
will lead the holon propagator to develop local criticality
\begin{equation}\label{}
G_{\chi } (\bx ,\tau ) \sim \frac{1}{\tau^{\alpha }}\delta ({\bx })
\end{equation}
The spinon self energy is given by a similar convolution of the holon
with the conduction electron, 
\[
\Sigma_{b } (\bx,\tau)= \frac{2S}{N}G^{{(0)}}_c(\bx,\tau)G_{\chi }
(\bx,\tau) 
\]
and the singular part of this self-energy will also be local
$\Sigma_{b} (\bx ,\tau )\sim \delta (\bx ) \frac{1}{\tau^{1+\alpha
}}$.  When this is combined with the effects of spinon pairing, it
leads to a spinon propagator of the form
\begin{equation}\label{}
G_{b} ({\bf   q } ,\omega )
\sim \frac{1}
{
({\bf   q } \pm {\bf Q}/2)^{2}+ \omega^{2\alpha }
}
\end{equation}
Thus although the spinons are not local in character, their singular
critical behavior will be  independent of momentum - a key element of
the observed local criticality.

At this stage, it remains to be seen whether
practical solution of the large $N$ equations can in practice furnish
this class of solutions, but the above arguments serve to illustrate
how the break-up of the
Kondo pseudo-potential and the reversion of the nominal valence 
real valence has the potential to drive locally quantum criticality.
The further investigation of these ideas is an ongoing element of our
current research. 



\section{Acknowledgement}
EL was supported in part by the Israeli ministry of 
absorption.
PC is supported by the National Science Foundation grant
NSF DMR-0605935.

\end{document}